\documentstyle[psfig]{mn}

\input{epsf}

\def\lsim{\mathrel{\hbox{\rlap{\hbox{\lower4pt\hbox{$\sim$}}}\hbox{$<$}}}}
\def\gsim{\mathrel{\hbox{\rlap{\hbox{\lower4pt\hbox{$\sim$}}}\hbox{$>$}}}}

\def\and   {\rm {et al.} \rm}  
\def\etal  {\rm {et al.} \rm}
\def\rmd {\rm d}

\begin{document}

\title[Cluster correlations in redshift space]
{Cluster correlations in redshift space}

\author[N. D. Padilla \& C. M. Baugh]
{N. D. Padilla$^1$ and C. M. Baugh$^2$
\\
1.IATE, Observatorio Astron\'omico de C\'ordoba, Laprida 854, 5000, C\'ordoba,
Argentina\\
2.Department of Physics, Science Laboratories, South Road, Durham DH1 3LE\\
\\
}

\maketitle 
 
\begin{abstract}
We test an analytic model for the two-point correlations of galaxy clusters 
in redshift space using the Hubble Volume N-body simulations.  
The correlation function of clusters shows {\it no} enhancement along 
the line of sight, due to the lack of any virialised structures in 
the cluster distribution.
However, the distortion of the clustering pattern due to coherent 
bulk motions is clearly visible.
The distribution of cluster peculiar motions is well described by 
a Gaussian, except in the extreme high velocity tails. 
The simulations produce a small but significant number of clusters 
with large peculiar motions.
The form of the redshift space power spectrum is strongly influenced 
by errors in measured cluster redshifts in extant surveys. 
When these errors are taken into account, the model reproduces the 
power spectrum recovered from the simulation 
to an accuracy of $15\%$ or better over a decade in wavenumber.
We compare our analytic predictions with the power spectrum measured 
from the APM cluster redshift survey. 
The cluster power spectrum constrains the amplitude of density fluctuations, 
as measured by the linear {\it rms} variance in spheres of radius 
$8 h^{-1}$Mpc, denoted by $\sigma_{8}$. 
When combined with the constraints on $\sigma_{8}$ and the density 
parameter $\Omega$ derived from the local abundance of clusters, 
we find a best fitting cold dark matter model with $\sigma_{8} 
\approx  1.25$ and $\Omega \approx 0.2$, for a power spectrum  
shape that matches that measured for galaxies. 
However, for the best fitting value of $\Omega$ and given 
the value of Hubble's constant from recent measurements, the assumed 
shape of the power spectrum is incompatible with the most readily 
motivated predictions from the cold dark matter paradigm.
\end{abstract}


\section{Introduction} 

Rich clusters of galaxies are unique tracers of the large scale 
structure of the Universe. This is due to a number of reasons. 
First, clusters are the most massive virialised systems in 
place at any given epoch, occupying a special place at the head of 
the structure hierarchy. 
Second, although rich clusters are rare objects they are also bright, 
containing many luminous galaxies and emitting copious 
amounts of X-rays. 
Third, 
a cosmologically interesting 
volume of the universe can be surveyed much more rapidly using 
clusters than with galaxies, as a result of the huge difference 
in the number of redshifts that are required to be taken. 
Finally, the key reason for the utility of clusters is that it is 
far easier to interpret the observed properties of the cluster 
distribution in the context of theoretical models than is the case 
for galaxies.

There are three main statistical properties of the cluster distribution 
that have been used to place constraints upon the parameters of 
structure formation models. 
The first of these is the local abundance of rich clusters, which, as 
a consequence of their rarity, is extremely sensitive to the amplitude 
of density fluctuations on a scale that encloses the mass of a 
typical rich cluster before collapse (White, Efstathiou \& Frenk 1993). 
The rate at which the abundance of clusters evolves with redshift further 
constrains the mass density of the universe (Eke, Cole, Frenk \& Henry 1998; 
Blanchard \etal 1998).
The second test is the distribution of cluster peculiar motions, which 
depends on the value of the mass density parameter $\Omega$, for models 
with similar mass fluctuation amplitudes (Croft \& Efstathiou 1994b; 
Bahcall, Cen \& Gramann 1994).
Lastly, the spatial correlations of clusters constrain both the 
shape and amplitude of the power spectrum of the underlying dark matter 
(Mo, Jing \& White 1996). 

The latter of these, the clustering of clusters, is by far the best studied 
and, at the same time, the most controversial. 
The first measurements of the spatial two-point correlation function 
of Abell clusters demonstrated a much stronger clustering amplitude 
than that found for galaxies (Bahcall \& Soneira 1983; 
Klypin \& Kopylov 1983).
Moreover, the clustering amplitude was found to increase significantly as the 
cluster number density decreased (Bahcall \& West 1992).
However, the exact correlation amplitude of  
clusters remains the subject of intense debate 
(Efstathiou \etal 1992; Miller \etal 1999). 
The early redshift surveys drawn from the Abell catalogue showed a significant 
enhancement of the clustering signal along the line of sight 
(Bahcall \& Soneira 1983; Postman, Huchra \& Geller 1992).
Bahcall, Soneira \& Burgett (1986) found that the clustering in the 
line of sight direction is consistent with a ``velocity broadening'' of 
$2000{\rm kms}^{-1}$, which they interpreted as arising from a combination 
of peculiar motions and geometrical distortions of superclusters.  

Confronted by these results, several authors have argued
that the Abell catalogue is afflicted by the superposition of 
clusters and that the clustering signal along the line of sight 
is artificial (Sutherland 1988; Sutherland \& Efstathiou 1991; 
see also Lucey 1983).
This prompted the construction of more objectively defined cluster 
catalogues drawn from machine-scanned survey plates with better 
calibrated photometry (APM: Dalton \etal 1992, 1994, 1997; 
Cosmos: Lumsden \etal 1992).
The typical radius used to define clusters in the machine 
based catalogues is significantly smaller than that used by Abell, 
reducing the enhancement of cluster richness by projection effects. 
The clustering signal found in these more recent cluster redshift surveys 
does not display large enhancements along the line of sight; furthermore, 
the trend of increasing correlation amplitude with decreasing space 
density of clusters is weaker than that found for Abell clusters 
(Croft \etal 1997). A similar dependence of 
correlation length on cluster space 
density is found for the X-ray Bright Abell Cluster Sample, 
which is much less susceptible to line of sight projection 
effects than the optically selected  Abell catalogue  
(Abadi, Lambas \& Muriel 1998, Borgani et al. 1999).  
This result was recently confirmed using the REFLEX survey in
a work by Collins et al. (2000), who found that there is no  significant
dependence of the clustering amplitud on the limiting flux, or
equivalently on the space density of galaxy clusters.

Miller \etal (1999) counter these objections by 
pointing out that the early redshift surveys of Abell clusters 
contained large fractions of low richness clusters 
(Abell richness class $R=0$), that were not intended to form complete samples 
for use in statistical analyses. 
Furthermore, many cluster positions were determined by a single galaxy 
redshift. 
Miller \etal (1999) present the clustering analysis of a new redshift 
survey of Abell clusters with richness $R \ge 1$, and with the majority of 
cluster positions determined using several galaxy redshifts. 
The clustering signal along the line of sight is greatly reduced in the 
new redshift surveys compared with the Bahcall \& Soneira (1983) 
results, and is comparable to the amount of distortion of the 
clustering pattern found for APM clusters (see Fig 5 of Miller \etal 1999).
The anisotropy is further reduced after the orientation of two 
superclusters that are elongated along the line of 
sight is changed. 
Peacock \& West (1992) also found that restricting attention to higher 
richness Abell clusters removed the strong radial anisotropy seen 
in the clustering measured in the earlier surveys.

It is clearly important to establish exactly what influence peculiar  
motions have on the inferred spatial distribution of rich clusters.  
In most previous theoretical studies, redshift space distortions 
have either been ignored or treated in a very approximate fashion. 
Quite often, redshift space distortions are modelled by simply assuming 
a boost to the clustering signal measured in real space, as predicted 
by Kaiser (1987).
This effect arises from coherent flows on large scales 
where linear perturbation theory is applicable. However, if 
the peculiar motions of clusters are significant, then a damping of 
the clustering signal is expected on small scales. 
The transition between these two extreme types of behaviour needs 
to be modelled.

Computer simulations of structure formation through the gravitational 
amplification of small primordial fluctuations have been used 
extensively to model the spatial distribution of clusters 
(e.g. White \etal 1987; Bahcall \& Cen 1992; Croft \& Efstathiou 1994a; 
Watanabe, Matsubara \& Suto 1994; Eke \etal 1996a).  
These early studies do not reach a consensus on the predicted 
clustering of clusters in cold dark matter cosmologies. 
Part of the reason for this discrepancy is due differences in the way in which 
clusters are identified in the simulations (Eke \etal 1996a). 
A further issue is the relatively small simulation volumes used and 
the small numbers of clusters analysed.

Recently, it has become possible to simulate much larger volumes than 
were used in these earlier studies, with sufficient resolution to 
allow the reliable extraction of massive dark matter haloes that 
can be identified as rich clusters (Governato \etal 1999; 
Colberg \etal 2000).  
In this paper, we analyse the redshift space clustering of massive 
dark matter haloes in the Hubble Volume simulations, the largest  
cosmological simulations to date, which are described in Section \ref{s:nbody}.
This extends the comparison carried out by Colberg \etal (2000), who 
measured clustering in real space and compared the results with the 
predictions of analytic models. We outline the analytic model that 
we employ to predict clustering in redshift space in Sections \ref{s:real} 
and \ref{s:pk}. The distribution of cluster peculiar velocities in the 
simulations, an important ingredient of the analytic model for redshift 
space clustering, is analysed in Section \ref{s:pec}. 
The predictions of the analytic model are confronted with measurements 
of clustering in the APM Cluster redshift survey in Section \ref{s:res}. 
We compare to cluster power spectrum data directly rather than to 
estimates of the correlation length. 
This avoids uncertainties introduced by the method used to 
derive a correlation length from the data. Moreover, the cluster power 
spectrum on large scales has the same shape as the power spectrum of the 
dark matter, as demonstrated in real space by Colberg \etal (2000),
and, as we show in Section \ref{s:pk}, is also the case under certain 
conditions in redshift space. 
We discuss our results and the constraints on cosmological parameters 
from the power spectrum of clusters in Section \ref{s:end}.

\section{Theoretical predictions for the cluster power spectrum}
\label{s:theory}

\subsection{N-body simulations}
\label{s:nbody}

We compare analytic predictions of the statistical properties of cluster 
samples with measurements made from the Virgo Consortium's 
``Hubble Volume'' simulations. 
The simulations follow the evolution of Cold Dark Matter density fluctuations 
in two cosmologies: $\tau$CDM (with cosmological 
parameters $\Omega = 1$, a power spectrum shape 
defined by $\Gamma = 0.21$, following the parameterisation given by 
Efstathiou, Bond \& White (1992), and a {\it rms} linear variance 
on a scale of $8h^{-1}$Mpc of $\sigma_{8} = 0.6$) and $\Lambda$CDM 
(with $\Omega_{0} = 0.3$, a cosmological constant 
$\Lambda_{0}c^{2}/(3H_{0}^{2}) = 0.7$, a power spectrum described 
by an effective shape parameter of $\Gamma = 0.17$ 
and $\sigma_{8} = 0.9$). The huge volume of the simulations 
($8h^{-3} {\rm Gpc}^{3}$ for $\tau$CDM and $27h^{-3} {\rm Gpc}^{3}$ for 
$\Lambda$CDM) and the large number of particles employed ($10^{9}$) 
allow cluster statistics to be studied with unprecedented accuracy 
(Colberg \etal 2000; Jenkins \etal 2001).
Dark matter haloes are identified using a friends-of-friends algorithm 
with a standard linking length (see Jenkins \etal 2001). 
The halo peculiar velocity is the peculiar motion of the centre of mass.

\subsection{Clustering in real space} 
\label{s:real}

In this section we review the formalism employed to model 
the power spectrum of clusters using positions measured in real 
space, i.e. ignoring any distortion to the power spectrum  
arising from the peculiar motions of clusters. 
For further details, we refer the reader to the 
more complete discussions of this framework given, for example, 
by Mo \& White (1996), Mo, Jing \& White (1996), Borgani \etal (1997),  
Colberg et al (2000) and Moscardini et al (2000). 
Moscardini \etal (2000) also consider the evolution of clustering 
along the observer's past light cone, an effect that we shall ignore for 
the relatively shallow observational sample studied in this paper.

Throughout this paper, we consider cluster samples that are defined by 
a characteristic spatial separation, $d_c$, or equivalently, by a 
space density $n$, where $d_{c} = 1/n^{1/3}$. Such a sample is 
constructed by first ranking the clusters in order of mass, 
and then, starting from the most massive cluster, including 
progressively less massive clusters until the required space density 
is achieved. 

We assume that on large scales, the real space power spectrum 
of the cluster sample, $P_{c}(k)$, can be related to the 
power spectrum of the underlying dark matter distribution, $P(k)$, 
by an effective bias factor, $b_{\rm eff}$, that is independent 
of scale:  

\begin{equation}
P_{c}(k) = b^{2}_{\rm eff} P(k). 
\label{eq:bias}
\end{equation}
Colberg et al (2000) use the Hubble Volume N-body simulations to 
demonstrate that this is an excellent approximation over a decade 
in wavenumber, $ 0.01 h {\rm Mpc}^{-1} < k <  0.1 h {\rm Mpc}^{-1}$.
The task of computing the real space power spectrum of clusters can 
therefore be broken down into two steps: (i) The calculation of 
the appropriate power spectrum of the dark matter distribution, $P(k)$, 
taking into account non-linear evolution of density fluctuations. 
(ii) The computation of the linear bias factor, $b_{\rm eff}$, for clusters of 
a given abundance.

The first stage in the calculation is carried out using the prescription 
for transforming a linear theory power spectrum into a non-linear power 
spectrum described by Peacock \& Dodds (1996). The transformation depends 
upon the cosmological parameters $\Omega$ and $\Lambda$, and upon the 
epoch or normalisation of the linear theory power spectrum. The formula 
given by Peacock \& Dodds agrees well with the non-linear evolution 
found in N-body simulations (Jenkins \etal 1998).

The effective bias is computed by taking a weighted average of 
the bias, $b(M)$, for haloes of mass $M$ over the cluster sample 
under consideration

\begin{equation}
b_{\rm eff} = \frac{ \int_{M_{\rm lim}}^{\infty} b(M) \frac{{\rm d} n(M)}{{\rm d}M} {\rm d}M}
{\int_{M_{\rm lim}}^{\infty}\frac{{\rm d}n(M)}{{\rm d} M} {\rm d} M}, 
\end{equation}
where $M_{\rm lim}$ is the lower mass limit that defines the sample 
and ${\rm d}n/{\rmd}M$ is the space density of halos in the 
mass interval $M$ to $M+\delta M$ 
(e.g. Mo \& White 1996; Governato \etal 1999).
We adopt the analytic form for the mass function of dark matter 
haloes proposed by Sheth, Mo \& Tormen (2001 - hereafter SMT). 
These authors put forward a modification to the theory of Press \& 
Schechter (1974) in which the collapse of dark matter haloes is 
followed using an ellipsoidal rather than spherical model. 
The SMT mass function agrees well with the results of 
N-body simulations, although the most significant improvements over 
Press-Schechter theory are realised for lower mass haloes than we consider 
in this paper (SMT; Jenkins \etal 2001). 
Following the theory developed by Mo \& White (1996), SMT also derive 
an expression for the bias factor of dark matter haloes (their equation 8) 
which we adopt in our calculations.

As reported by Colberg \etal (2000), the SMT formulae for the mass function 
and halo bias factor predict an effective bias that is in good 
agreement with the results obtained from the Hubble Volume simulations. 
The analytic predictions for the real space power spectrum of clusters 
with $d_{c}=30.9h^{-1}$Mpc are shown by the solid lines in the upper 
panels of Fig. \ref{fig:pks}. 
The discrepancy is largest for the $\Lambda$CDM model, in which case the 
analytic prediction for the real space power spectrum is $12\%$ higher 
than the measurement from the simulation.

\subsection{Cluster peculiar velocities}
\label{s:pec}

\begin{figure*}
{\epsfxsize=18.truecm \epsfysize=15.truecm 
\epsfbox[40 210 570 630]{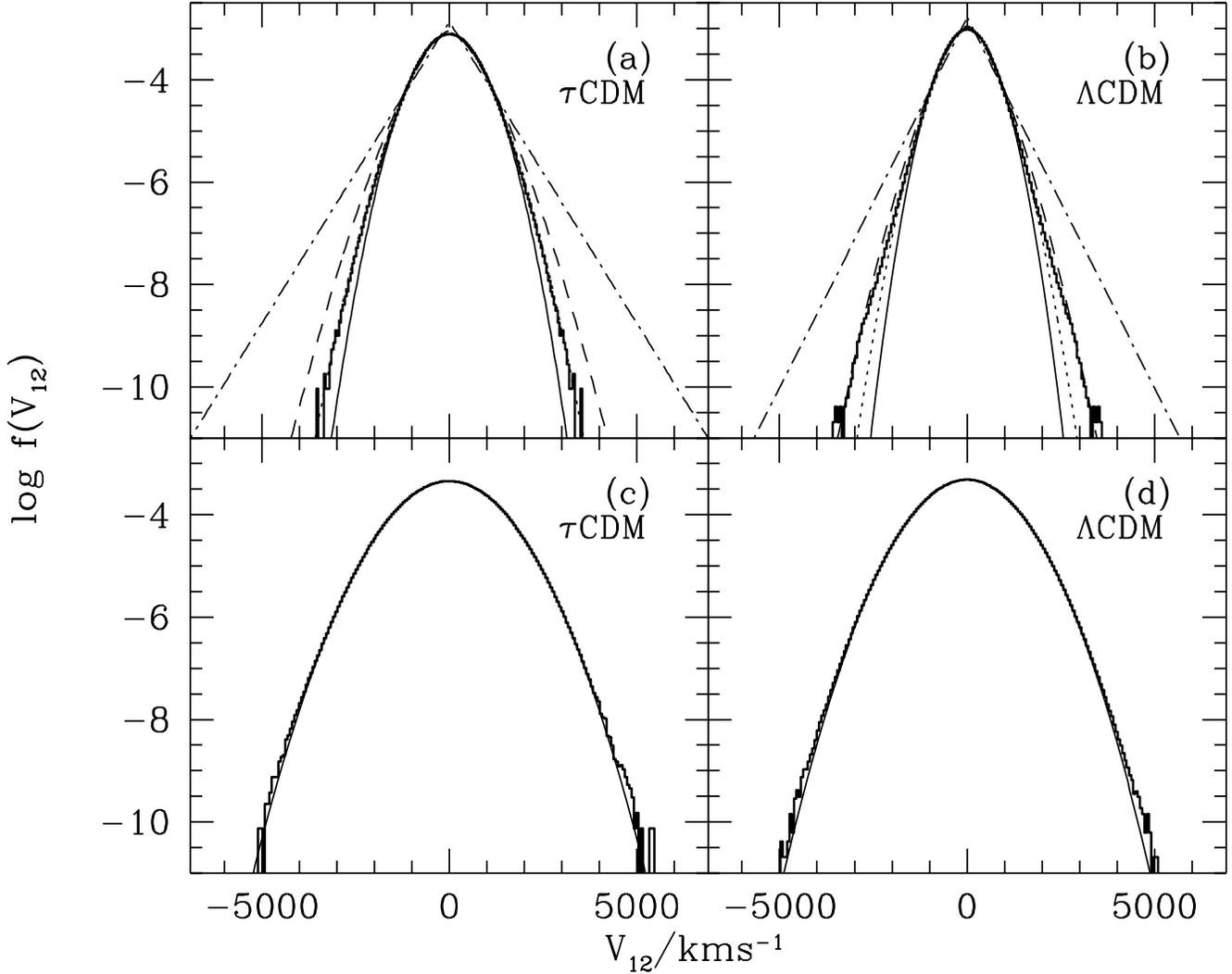}}
\caption{
The histograms show the distribution of line of sight pairwise peculiar 
velocities of clusters with $d_c=30.9 h^{-1}$Mpc in the Hubble 
Volume simulations: (a) shows the results for $\tau$CDM and 
(b) for $\Lambda$CDM, as indicated in the legend. 
The lower panels show the distributions when a single cluster {\it rms} 
redshift error of $\delta v  = 500 {\rm kms}^{-1}$ is included. 
The smooth curves show theoretical distributions plotted with the 
same variance in pairwise peculiar velocity that is measured from the 
simulations. In (a) and (b), solid lines show a Gaussian 
distribution, dashed lines show an exponential distribution in 
$|v_{12}|^{3/2}$, dotted lines are for an exponential distribution 
in $|v_{12}|^{7/4}$, and the 
dot-dashed lines show an exponential distribution in $|v_{12}|$. 
In the lower panels, the variance includes the redshift error described 
above, added in quadrature to the variance measured in the simulation.
In (c) and (d) only the Gaussian distribution is plotted.
}
\label{fig:f_v}
\end{figure*}

The gravitationally induced peculiar motions of clusters distort the 
pattern of clustering if cluster redshifts are used to infer their 
spatial distribution. The distribution of cluster peculiar motions 
is therefore a key ingredient in the theoretical prediction of clustering 
in redshift space, as discussed in the next section.

In Fig. \ref{fig:f_v}(a) and (b), the histograms show 
the distribution of line of sight 
pairwise peculiar velocities, $f(v_{12})$, measured for clusters 
in the Hubble Volume simulations with $d_c=30.9 h^{-1}$Mpc; 
(a) shows the distribution for the $\tau$CDM simulation and 
(b) shows the results for $\Lambda$CDM.
The smooth curves show various analytic distributions plotted with the 
pairwise velocity dispersion that is measured in the simulations 
($\sqrt{< v^{2}_{12}>}=532 {\rm kms}^{-1}$ for clusters in the $\tau$CDM 
simulation and $\sqrt{<v^{2}_{12}>}=434 {\rm kms}^{-1}$ for $\Lambda$CDM);
the solid line shows a Gaussian distribution, 
the dashed line shows an exponential distribution in $|v_{12}|^{3/2}$, 
the dotted line shows an exponential distribution in $|v_{12}|^{7/4}$ 
and the dot-dashed line shows an exponential distribution in $|v_{12}|$ 
(full details of the functional forms of these distributions may be found 
in Padilla \& Baugh 2001). 
The bulk of the distribution of cluster peculiar velocities 
is adequately described by a Gaussian. 
However, Fig. \ref{fig:f_v} illustrates that   
this is not the case for the high velocity tails of the distributions.
Moreover, the best fitting distribution is not the same in each simulation. 
Padilla \& Baugh (2001) show that the shape of the high velocity 
tail depends upon the degree of non-linear evolution of the density 
fluctuations. 
The distribution of peculiar velocities is therefore of interest in itself, 
being sensitive to the cosmological parameters $\Omega$ and 
$\sigma_{8}$, and to the power spectrum of density fluctuations 
(Croft \& Efstathiou 1994b). These issues are explored in more 
detail using the Hubble Volume simulations and other N-body 
simulations by Padilla \& Baugh (2001).

The histograms in the lower panels of Fig. \ref{fig:f_v} 
show the distribution of peculiar velocities after incorporating 
a single cluster {\it rms} redshift measurement error of 
$\delta v \approx 500 {\rm kms}^{-1}$, 
appropriate for the Abell and APM cluster redshift surveys 
(Efstathiou \etal 1992).
The resulting distributions are much broader and closer to 
Gaussian, as shown by the corresponding solid lines; in these cases 
the variance is given by the sum in quadrature of the variance 
measured in the simulation 
and the {\it rms} redshift error stated above. The redshift error dominates 
over the variance expected for gravitationally induced peculiar velocities  
for the models that we consider in this paper. 
We therefore make the approximation in subsequent calculations when 
comparing to APM data that the 
variance in the pairwise peculiar velocity, including errors, is 
fixed at $\sqrt{<v^{2}_{12}>} \sim 850 {\rm kms}^{-1}$, regardless 
of the model being studied.
This in turn implies a single particle {\it rms} peculiar velocity of 
$\sigma = \sqrt{<v^{2}_{12}>}/\sqrt{2} \approx 600 {\rm kms}^{-1}$.

\subsection{Clustering in redshift space - power spectrum}
\label{s:pk}
\subsubsection{Analytic model}

The distortion of the power spectrum measured in redshift space 
generally displays two forms. 
On large scales, the amplitude of the power spectrum is boosted due to 
coherent inflows into overdense regions and outflows from underdense 
volumes (Kaiser 1987). On small scales, randomised motions within 
virialised structures cause a damping of power (e.g. Peacock 1999).
Simple models have been developed that describe the transition between 
this large and small scale behaviour (Peacock \& Dodds 1994, 
Cole, Fisher \& Weinberg 1995). 
These schemes have been shown to work reasonably well for dark 
matter in N-body simulations (Cole, Fisher \& Weinberg 1995; 
Hoyle \etal 1999). 
In this section we test whether such models provide an accurate description 
of the redshift space power spectrum of clusters of galaxies; 
this is necessary as we do not expect to find virialised structures 
in the cluster distribution.

For scales that are still evolving according to linear perturbation theory, 
the power spectrum in redshift space is given by 

\begin{equation}
P^{s}_{c}(k,\mu) = P_{c}(k) \left( 1 + \beta \mu^{2} \right)^{2}, 
\label{eq:kaiser}
\end{equation}
where $P_{c}$ is the cluster power spectrum in real space, as defined by 
equation \ref{eq:bias}, $P^{s}_{c}(k)$ is the cluster power spectrum 
in redshift space, $\beta = f(\Omega)/b_{\rm eff}$ ($f(\Omega)$ is the 
logarithmic derivative of the fluctuation growth rate) and 
$\mu$ is the cosine of the angle between the wavevector $k$ and 
the line of sight (Kaiser 1987; see also the discussion of 
this result in Cole, 
Fisher \& Weinberg 1994).

Heuristic schemes have been put forward that extend this model for the 
redshift space power spectrum down to small scales to include the 
effects of a random velocity dispersion, under the assumption that the 
velocities are uncorrelated with the density field: 
\begin{equation}
P^{s}_{c}(k,\mu) = P_{c}(k) \left( 1 + \beta \mu^{2} \right)^{2} 
D(k\mu \sigma_{v}). 
\label{eq:pkmu}
\end{equation}
For the case of a Gaussian distributed velocity dispersion, 
\begin{equation}
D(k \mu \sigma_{v}) = \exp(-k^{2}\mu^{2}\sigma^{2}_{v}/2),
\end{equation}
whilst for an exponential distribution (Ballinger, Peacock \& Heavens 1996),   
\begin{equation}
D(k \mu \sigma_{v}) = \frac{1}{1 + \left(k\mu\sigma_{v}\right)^{2}/2}. 
\end{equation}
The spherically averaged form of equation \ref{eq:pkmu} for a 
Gaussian velocity dispersion is given by Peacock \& Dodds (1994): 
\begin{equation}
P^{s}_{cl}(k) = G(\beta, y) P_{cl}(k),
\end{equation}
where the function $G(\beta, y)$ is given by
\begin{eqnarray}
G(\beta, y) & = \frac{\sqrt{\pi}}{8} \frac { {\rm{erf}}(y)}{y^5}
              [ 3 \beta^2 + 4 \beta y^2 + 4 y^4] \nonumber \\
            &  - \frac{\exp(-y^2)}{4 y^4}
              [\beta^2 (3+2 y^2) + 4 \beta y^2],
\label{eq:G}
\end{eqnarray}
where $\beta = f(\Omega)/b_{\rm eff}$ and 
$y=k\sigma_{v}/
(100{\rm kms}^{-1}{\rm Mpc}^{-1})$.
Errors in the determination of the cluster redshifts can be incorporated 
into this model by adding the redshift error in quadrature to the {\it rms} 
peculiar velocity to redefine $\sigma_v$.
On the scales that we consider, there is effectively no difference in the 
distortion to the power spectrum when a Gaussian or exponential distribution 
of velocity dispersion is adopted (see Fig. 3 of 
Ballinger, Heavens \& Peacock 1996).

\subsubsection{Comparison with simulation results}

\begin{figure*}
{\epsfxsize=18.truecm \epsfysize=15.truecm 
\epsfbox[50 220 550 620]{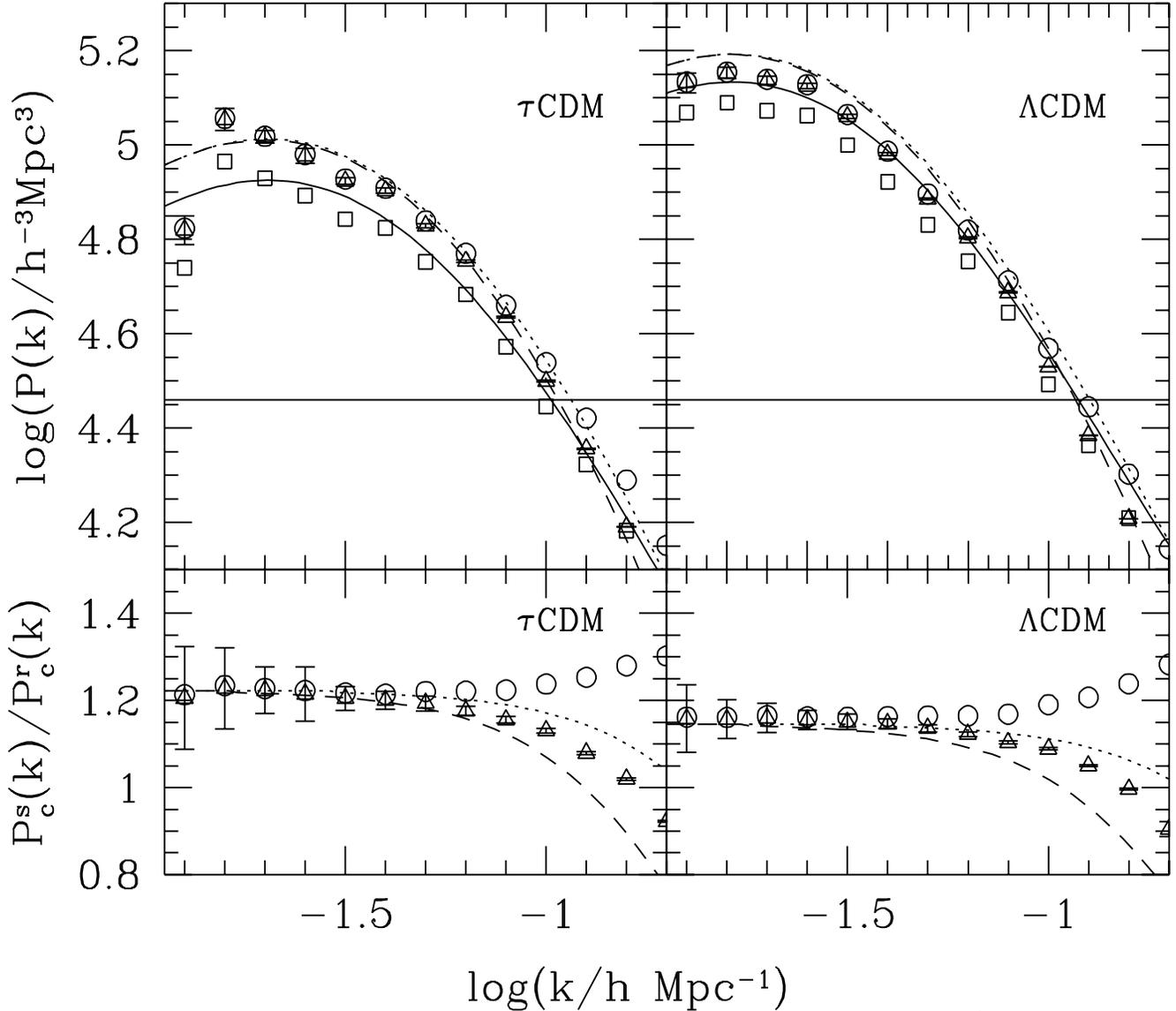}}
\caption{
The upper two panels show the real and redshift space power spectra 
for clusters with $d_{c}=30.9h^{-1}$Mpc; the left panels show the results for 
$\tau$CDM and the right panels show results for $\Lambda$CDM. 
The lines show analytic predictions 
and the symbols show measurements made directly from the simulations. 
Real space quantities are shown by solid lines and squares. 
The redshift space power spectrum measured in the simulations is shown 
by the circles.
The dashed lines and triangles show the redshift space power spectrum 
after incorporating a redshift error into the determination of cluster 
positions.
The lower panels show the ratio between redshift space and real space 
power spectra as a function of wavenumber. Again, points show the 
ratio for the measurements in the simulation. The circles show the ratio 
when only peculiar motions are considered, the triangles show the ratio 
when redshift errors are also included.
The dashed line shows the 
analytic prediction of the ratio, when cluster redshift errors are included 
and should be compared with the triangles.
The error bars are set by the number of modes per bin in wavenumber in the 
simulation.
}
\label{fig:pks}
\end{figure*}

The analytic prediction for the spherically averaged cluster power spectrum 
in redshift space is compared to the measurements from the Hubble 
Volume simulations in Fig. \ref{fig:pks}. 
In the simulations, the clustering pattern in redshift space 
is obtained by displacing clusters along the x-axis by an amount 
$\delta x = v_{x}/H_{0}$, where $v_{x}$ is the x-component of a cluster's 
peculiar motion.
The power spectrum of clusters in the Hubble Volume simulation 
is computed using the technique described in detail by Jenkins \etal (1998), 
and is equivalent to employing a very high resolution grid 
to perform the Fast Fourier Transform (FFT). 
Therefore, the scheme used to assign clusters to the FFT grid 
does not influence the recovered power spectrum.

The lower panels in Fig. \ref{fig:pks} show the ratio of the redshift 
space to real space power spectrum as a function of wavenumber. 
The dashed lines show the outcome of dividing the analytic prediction for 
the real space spectrum into the prediction for the redshift space 
power spectrum, calculated including the effects of cluster redshift errors.
At small wavenumbers (large scales) the ratios are in excellent agreement 
with the expectations from equation \ref{eq:kaiser}, which, using the 
approximation $f(\Omega) \approx \Omega^{0.6}$  predicts a ratio of $1.22$ 
for clusters in $\tau$CDM and $1.15$ in the $\Lambda$CDM simulation. 
Note the latter value changes by less than $1\%$ if the weak influence 
of a non-zero cosmological constant on $f(\Omega)$ is taken into 
account (Lahav \etal 1991). 
At high wavenumbers (small scales) we find that there is no damping of 
the power spectrum measured in redshift space when redshift errors are 
ignored.  
The analytic model reproduces the boost in power on large scales but 
predicts too much damping in the power on small scales.
If an error is included in the cluster redshifts, then the model 
predicts the same form of distortion found in the simulations.
The upper panels of Fig. \ref{fig:pks} show that in this case, 
the analytic model for the redshift space power spectrum 
agrees with the simulation results to $15\%$ or better over the 
wavenumber range $-2<\log(k/h{\rm Mpc}^{-1})< -1$.

\begin{figure}
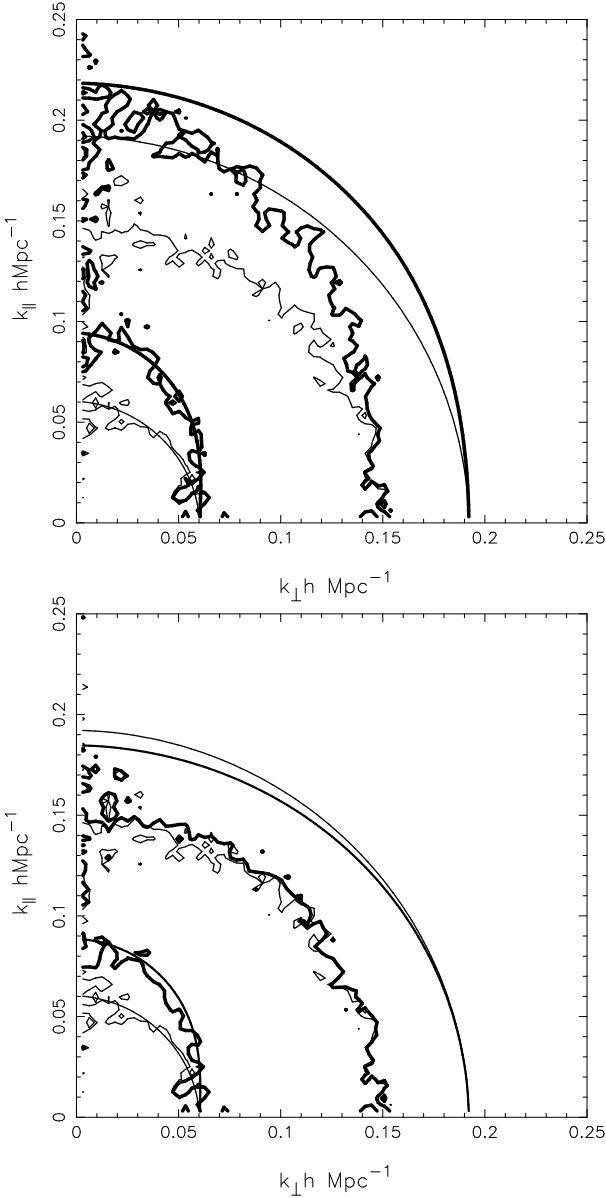

\begin{picture}(200,450)
\put(0,225){\psfig{file=pk2.tcdm.1.ps,width=8cm,height=8cm,angle=-90}}
\put(0,0){\psfig{file=pk2.tcdm.2.ps,width=8cm,height=8cm,angle=-90}}
\end{picture}
\caption{
The power spectrum as a function of wavenumber parallel $k_{||}$ and 
perpendicular $k_{\perp}$ to the line of sight. The smooth lines show 
theoretical predictions, the other lines show measurements from the 
$\tau$CDM Hubble Volume simulation. The light lines in both panels 
show the power spectrum in real space. In the upper panel, the heavy 
lines show the power spectrum when the peculiar motions of clusters 
are taken into account. In the lower panel, redshift errors in the cluster 
positions are included. 
The inner most set of contours show the wavenumbers for which the power 
is $4.9 \times 10^{4} h^{-3}{\rm Mpc}^{3}$; the outermost set show 
$P(k_{\perp},k_{||})= 1.2 \times 10^{4} h^{-3}{\rm Mpc}^{3}$
}
\label{fig:pkmu}
\end{figure}

Further insight into the form of the distortion of the redshift 
space power spectrum caused by cluster motions can be obtained 
by plotting the power spectrum as a function of wavenumber perpendicular 
($k_{\perp}$) and parallel ($k_{||}$) to the line of sight, 
$P(k_{\perp}, k_{||})$, where 
$k = \sqrt{k^{2}_{\perp} + k^{2}_{||}}$ and $\mu = k_{||}/k$.  
We plot $P(k_{\perp}, k_{||})$ for $d_c = 30.9 h^{-1}{\rm Mpc}$ 
clusters in the $\tau$CDM Hubble Volume simulation in Fig. \ref{fig:pkmu}. 
The smooth lines show analytic predictions. The light solid lines 
show the real space power spectrum.
In the upper panel, the heavy solid lines show the redshift 
space power spectrum.  
In the lower panel the heavy lines show the power spectrum in redshift 
space including the effects of redshift errors in the determination 
of cluster positions.  
Two contour levels are plotted; the innermost sets of contours 
show $P(k_{\perp}, k_{||})=4.9 \times 10^{4} h^{-3}{\rm Mpc}^{3}$ 
and the outermost set show $P(k_{\perp}, k_{||})=1.2 \times 10^{4} 
h^{-3}{\rm Mpc}^{3}$. 
On large scales, the analytic predictions are in excellent agreement  with the 
measurements from the simulations, in real space and in redshift space. 
In redshift space, the power is enhanced on large scales in the 
$k_{||}$ direction, displacing the contour of fixed power to higher 
wavenumbers. On intermediate scales, the distortion of the power 
spectrum is extremely sensitive to how well cluster redshifts are 
measured. The magnitude of the error estimated in the redshifts of 
APM and Abell survey clusters dominates over the distortion due to 
the peculiar motions of the clusters. 
Therefore, on these scales, the boost in power given by equation 
\ref{eq:kaiser} is a poor description of the power spectrum.
Qualitatively, the analytic model reproduces the form of the redshift 
space distortion measured in the simulation. However, due to a 
small discrepancy on intermediate scales between the predicted real 
space power spectrum and the measurement from the simulation 
(of around $10\%$ in the amplitude of $P(k)$), the contours do not 
coincide on this plot.

\begin{figure}
{\epsfxsize=8.1truecm \epsfysize=8.1truecm 
\epsfbox[30 150 550 690]{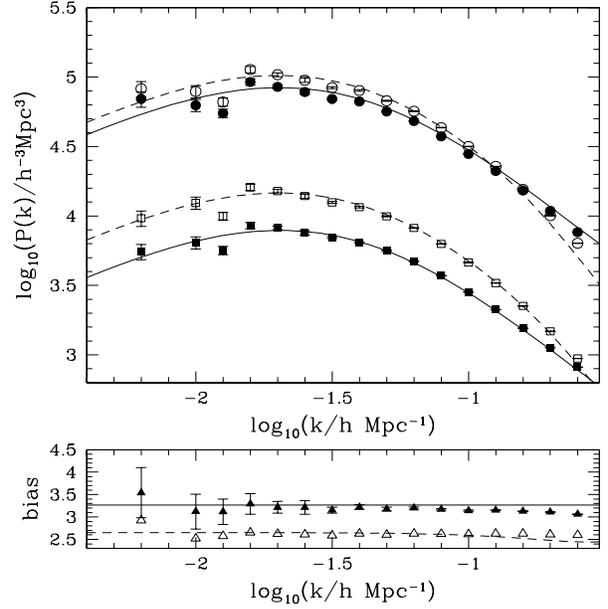}}
\caption{
A comparison of real and redshift space power spectra for 
$d_{c}=30.9h^{-1}$Mpc clusters and for dark matter in the 
$\tau$CDM simulation. Solid symbols and lines show real space 
quantities and open symbols and dashed lines show redshift 
space quantities. 
Squares show power spectra for the dark matter and circles show 
cluster power spectra.
Redshift errors have been included in the 
redshift space power spectrum of clusters. 
The solid lines in the upper panel show the fits to
the real-space CDM model for the mass and clusters,
and the dashed lines show the same fits for
redshift space.
The lower panel shows the effective bias obtained by taking 
the square root of the ratio between cluster and mass power spectra, 
both in real space (solid points and lines) and in redshift space 
(open points, dashed line).
}
\label{fig:pkb}
\end{figure}

Finally, we compare the shapes of the redshift space 
power spectrum of clusters and dark matter in the $\tau$CDM Hubble 
Volume simulation in Fig \ref{fig:pkb}.
The results for clusters include errors in the cluster 
redshift determination, as discussed above.
This plot illustrates that the redshift space power spectrum predicted by 
the model is in very good agreement with that measured for the dark matter.
The effective bias measured in redshift space, as deduced by taking the ratio 
of cluster and mass power spectra in redshift space, is somewhat lower 
than the bias measured in real space. 
However, the bias measured in redshift space is still independent of scale. 
A measurement of the cluster power spectrum on large scales in redshift 
space would therefore yield the shape of the mass power spectrum in 
redshift space.

\subsection{Clustering in redshift space - correlation function}
\label{s:clusz}

A key statistic for cluster samples is the two point correlation 
function $\xi$, measured as a function of cluster separation parallel 
to the line of sight, $\pi$, and perpendicular to the line of sight, $\sigma$.
Efstathiou \etal (1992) used this statistic to argue that cluster 
samples drawn from the Abell catalogue are contaminated by projection 
effects, leading to a spurious enhancement of the clustering signal 
along the line of sight. 
Previously, the two-point correlation function $\xi(\sigma,\pi)$ has 
been studied for more abundant clusters, using much smaller volume 
simulations than are considered in this paper (e.g. Eke \etal 1996a). 
The Hubble Volume simulations can be used to resolve once and for all 
the issue of exactly how much anisotropy is expected in the two 
point correlation function from redshift space distortions alone.

\begin{figure*}
\centerline{\psfig{file=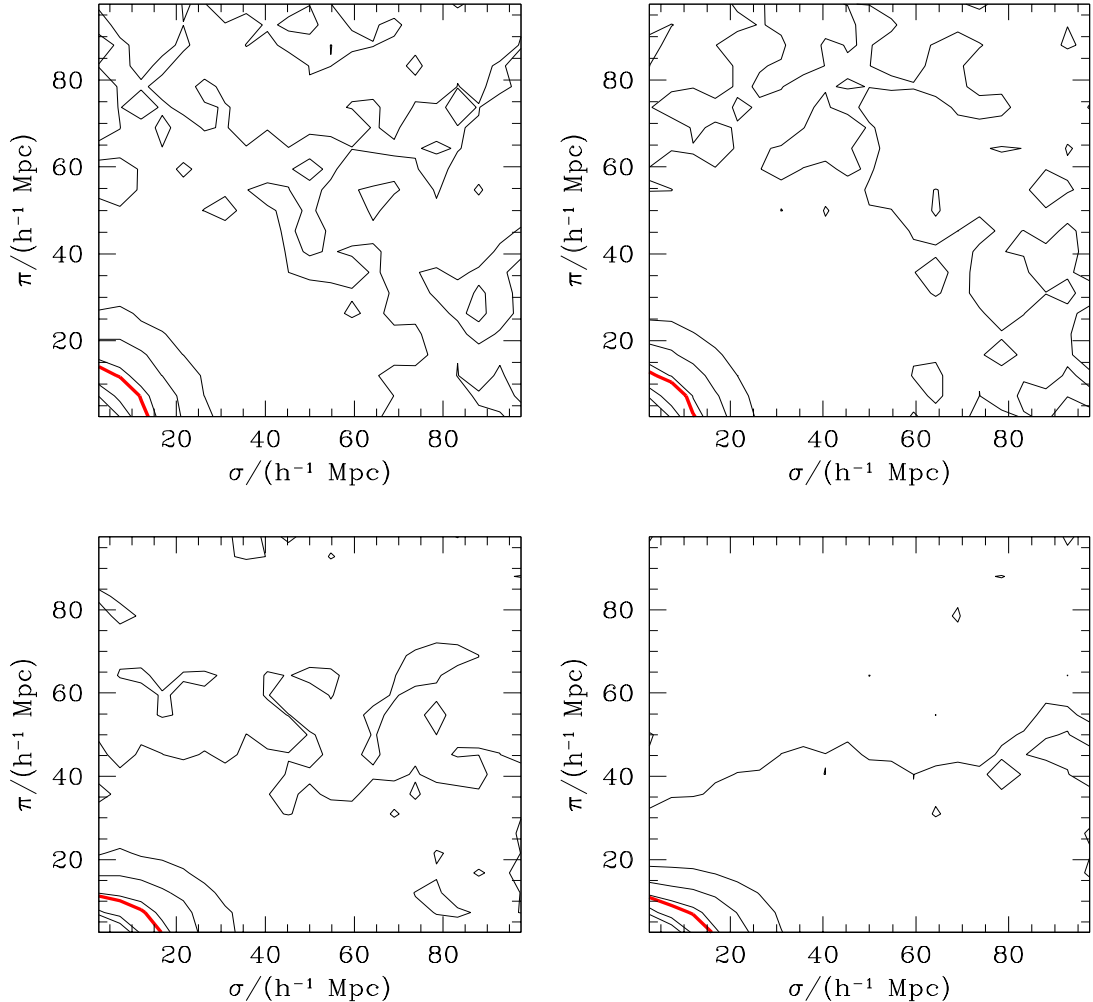,width=16cm,height=15cm}}
\caption{ 
The two-point correlation function $\xi(\sigma,\pi)$ plotted in 
bins parallel ($\pi$) and perpendicular ($\sigma$) to the line 
of sight for $d_{c}=30.9h^{-1}$Mpc clusters in the $\tau$CDM 
(left-hand panels) and $\Lambda$CDM (right-hand panels) Hubble 
Volume simulations.
The upper row shows the correlation function in real space and the lower 
row shows the redshift space correlation function. 
Contour levels are at $\xi=3,2,1,0.8,0.6,0.4,0.2$. The thick 
contour shows $\xi=1$.
}
\label{fig:sigpi1}
\end{figure*}

In Fig \ref{fig:sigpi1}, we show $\xi(\sigma,\pi)$ for clusters defined 
by $d_{c} = 30.9h^{-1}$Mpc in the Hubble Volume simulations. 
The left hand panels show correlation functions measured in the $\tau$CDM 
simulation and the right hand panels show results for $\Lambda$CDM.
The upper row of Fig. \ref{fig:sigpi1} shows the correlation function  
measured in real space and the lower row shows the redshift space correlation 
function. Note that in this Fig., we do not include any redshift errors. 
The contour levels are given in the figure caption; the thick contour 
shows $\xi=1$.
As expected, the contours do not show any distortion when the correlation 
function is measured in real space. 
However, when the effects of peculiar motions are included, we find an 
apparent enhancement in the clustering signal {\it perpendicular} to 
the line of sight.
This is a result of the flattening of the contours in the 
$\pi$ direction due to coherent flows in the cluster distribution.
There is no evidence for {\it any} enhancement of the clustering signal 
along the line of sight. 
As a test, we have also measured $\xi(\sigma,\pi)$ for the dark matter 
in the Hubble Volume simulations, and in this case we do find a strong 
enhancement of the clustering amplitude in the $\pi$ direction on 
small scales. 
The relatively large separation of particles on the initial grid in the 
simulations ($3h^{-1}$Mpc in the case of $\Lambda$CDM) is therefore not 
an issue.
The explanation of our result for clusters is that virialised structures 
have not had time to form in the cluster distribution.

\begin{figure}
\centerline{\psfig{file=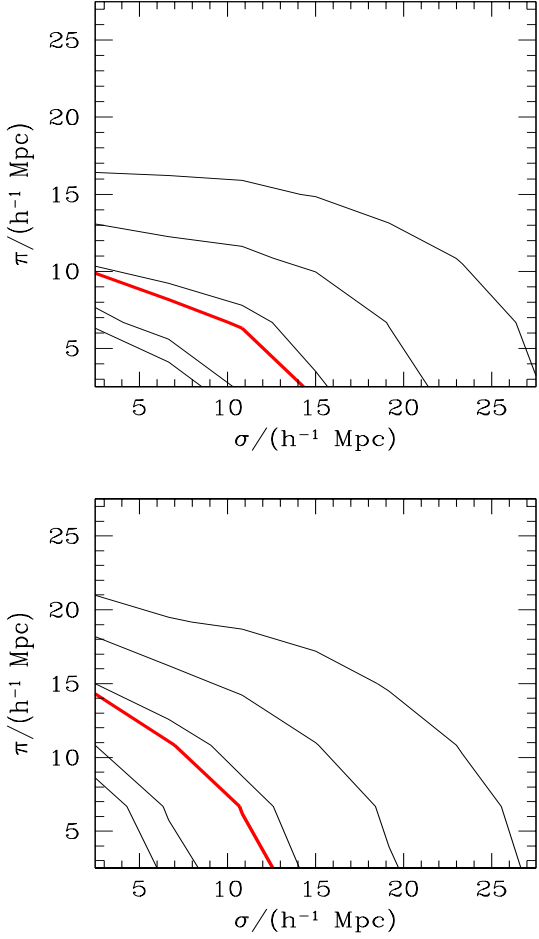,width=16cm,height=14cm}}
\caption{ 
The two-point correlation function measured in redshift space 
for clusters in the $\tau$CDM simulation. 
The same contour levels used in Fig. \ref{fig:sigpi1} are plotted, 
with the thick contours showing $\xi=1$. 
In panel (a), gravitationally induced peculiar motions are considered. 
In (b), a {\it rms} cluster redshift error of $500 {\rm kms}^{-1}$ is 
also included.
}  
\label{fig:sigpi2}
\end{figure}

The correlation function measured in the $\tau$CDM simulation is 
shown on an expanded scale in Fig. \ref{fig:sigpi2}. 
In the lower panel, we include a {\it rms} cluster redshift error 
of $500{\rm kms}^{-1}$, in addition to the peculiar motions. 
When cluster redshift errors are included, a clear boost is evident 
in the amplitude of clustering parallel to the line of sight. 
It is interesting to note that the flattening of the contours due to 
coherent flows is no longer apparent; the magnitude of the cluster redshift 
errors is sufficient to obscure this effect.
It can also be seen that the clustering amplitud along the $\sigma$
direction shortens noticeably.  This can be explained as follows:
the effect of the redshift errors
on the correlation function at fixed values of $\sigma$, will be that
of a smoothing function, which will take power from the scales 
corresponding to the higher correlations and transfer it to those
of lower correlation.  Therefore the value of $\xi(\sigma,\pi=0)$
will be smaller after the redshift errors are included, and the correlation
lenght in the direction perpendicular to the line of sight will be
shorter.

\section{Comparison of the theoretical model with data}
\label{s:res}

\subsection{Clustering data}
\label{s:data}

\begin{figure}
{\epsfxsize=8.1truecm \epsfysize=10.1truecm 
\epsfbox[40 150 520 700]{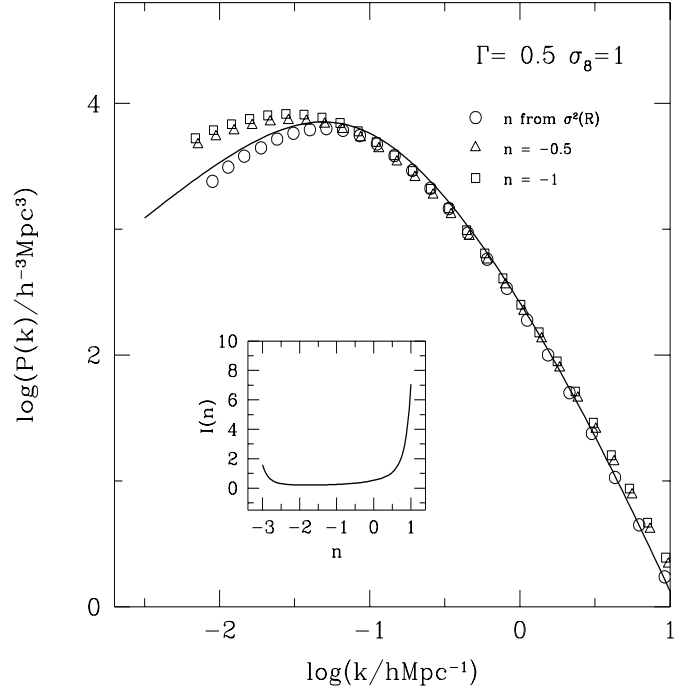}}
\caption{
Testing the recovery of the power spectrum from the variance. 
The solid line shows a linear theory CDM power spectrum with 
$\sigma_{8}=1$ and $\Gamma=0.5$. The points show the power spectra 
recovered by applying equations \ref{eq:var} and \ref{eq:keff} to the 
variance computed from the linear theory power spectrum (as 
given by equation \ref{eq:sigma}). The different symbols denote 
the results obtained for different assumptions about the slope 
of the power spectrum.
}
\label{fig:pktest}
\end{figure}

\begin{figure}
{\epsfxsize=8.1truecm \epsfysize=10.1truecm 
\epsfbox[40 150 520 700]{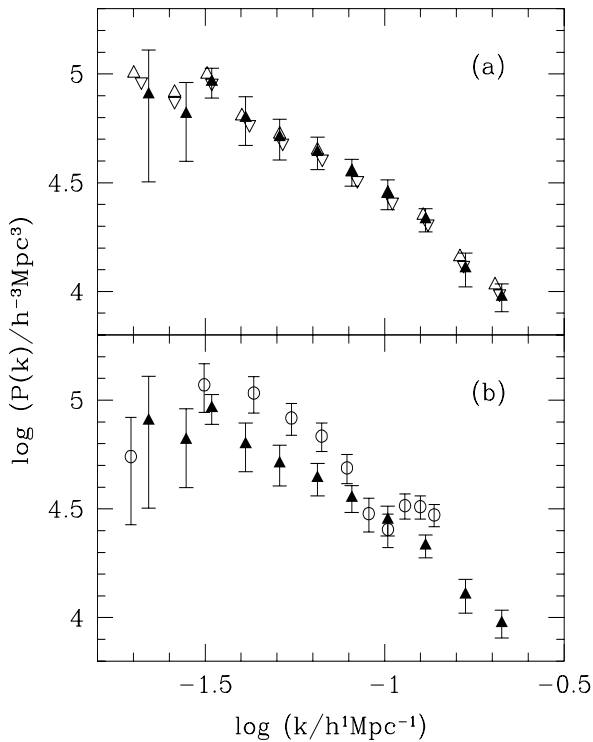}}
\caption{
(a) Estimates of the power spectrum of APM clusters inferred by applying  
equations \ref{eq:var} and \ref{eq:keff} to the variance of 
counts in cells measured by Gazta\~{n}aga \etal (1995). 
The different symbols show the results for different assumptions about the 
slope of the power spectrum: the solid triangles use the slope of the 
variance data directly, the upwards pointing triangles are for a fixed 
value of $n =-1$ and the downwards pointing triangles are for $n=-0.5$.
(b) A comparison of the power spectrum inferred using the counts-in-cells 
measurements of Gazta\~{n}aga \etal (1995) (filled triangles) with the 
direct measurement of the power spectrum of APM clusters by Tadros \etal 
(1998) (circles).
}
\label{fig:pkdata}
\end{figure}

In this paper we compare the predictions of CDM models with 
the power spectrum measured from the APM cluster redshift 
survey.
The data we consider are the variance of counts in cells 
measured by Gazta\~{n}aga, Croft \& Dalton (1995) and the power 
spectrum measured by Tadros, Efstathiou \& Dalton (1998).
Both measurements were made using sample B of the APM 
cluster redshift survey, as defined by Dalton \etal (1994).

Before confronting the model predictions with these data, we first 
compare the two measurements with one another. 
The variance of counts in cells is given by an integral over 
the power spectrum multiplied by the Fourier transform of the 
window used to smooth the density field (e.g. Peacock 1999) 
\begin{equation}
\sigma^{2}(R) = \frac{1}{2\pi^{2}} \int_{0}^{\infty} {\rm d} k k^{2} P(k) 
W^{2}(kR).
\label{eq:sigma}
\end{equation}
For a spherical top hat smoothing window of radius $R$, 
\begin{equation}
W(kR) = \frac{3}{(kR)^{3}} \left( \sin(kR) - kR \cos(kR) \right).
\end{equation}
The integral is reasonably sharply peaked around a characteristic 
wavenumber for a particular smoothing scale. 
Therefore, we make the approximation that the variance measured in a 
sphere of radius $R$ can be related to the power spectrum at a 
specified wavenumber (see Peacock 1991 who gives the result for 
a Gaussian smoothing window): 
\begin{equation}
\sigma^{2} (R) = \Delta^{2}(k_{\rm eff}),
\label{eq:var}
\end{equation}
where $\Delta^{2}(k) = k^{3}P(k)/2 \pi^{2}$. 
If we consider power law spectra, $P(k) \propto k^{n}$, 
then the effective wavenumber is defined as:

\begin{equation}
k_{\rm eff} = \frac{1}{R} \left[ 9 I(n)\right]^{1/(n+3)},
\label{eq:keff}
\end{equation}
Here $n$ is the logarithmic slope of the power spectrum at $k_{\rm eff}$, 
and the function $I(n)$ is defined by (see also Gazta\~{n}aga 1995):
\begin{equation}
I(n) = \int_{0}^{\infty} {\rm d} x x^{n-4} 
\left( \sin x - x \cos x\right)^{2}.
\label{eq:in}
\end{equation}
We demonstrate the accuracy of this transformation in Fig. \ref{fig:pktest}.
The points show the power spectra recovered by applying 
equations \ref{eq:var}, \ref{eq:keff} and \ref{eq:in} to the variance 
computed by integrating over the power spectrum shown by the solid line, 
which is our input spectrum 
(using equation \ref{eq:sigma}).
The different symbol types delineate the results obtained when different 
assumptions are made about the shape of the power spectrum. 
The most accurate answers are obtained when the shape of the power spectrum 
is inferred directly from the local slope of the variance. In this case, 
the recovered spectrum is at most $10\%$ below the correct value.
The location of the turnover is reproduced when fixed values of the spectral 
index $n$ are used; however, the amplitude of the recovered spectrum can 
differ from that of the true spectrum by a factor of two, on scales larger  
than the turnover.

In Fig \ref{fig:pkdata}(a), we show estimates of the power spectrum made 
by applying equations \ref{eq:var} and \ref{eq:keff} to the variance 
of counts in cells for APM clusters taken from Figure 3 of Gazta\~{n}aga 
\etal (1995). The different symbols show the results for different assumptions 
about the logarithmic slope of the power spectrum: the filled triangles 
show the power spectrum when the slope $n$ is estimated directly 
from the measured variance; the upwards pointing, open triangles 
show the results for a fixed value of $n=-1$  
and the downwards pointing triangles show the case where $n=-0.5$. 
The power spectrum recovered in this way from the variance 
is robust to reasonable variations in the value of the spectral 
index adopted in equation \ref{eq:keff}. 

We compare the power spectrum estimated from the variance of counts 
in cells, using the value of $n$ inferred from the data, with 
the power spectrum measured directly in redshift space by Tadros \etal (1998)
in Fig \ref{fig:pkdata}(b). 
Over most of the range considered, the Tadros \etal (1998) 
result is approximately 
$50\%$ higher in amplitude than that obtained by Gazta\~{n}aga \etal (1995)
Furthermore, though both measurements agree within the errors on the 
largest scales, the suggestion of a turnover seen in the Tadros \etal (1998)
power spectrum is not as apparent in the power spectrum inferred from the 
variance data.
In both papers the same sample of clusters was considered, with the 
same cuts applied in redshift. 
Both approaches make assumptions that could be partly responsible for 
the discrepancy.
Tadros \etal (1998) need to specify a radial weighting scheme to obtain a 
minimum variance estimate of the power spectrum. The weighting scheme 
depends upon the power spectrum, but in practice the recovered 
spectrum is fairly insensitive to the value chosen.
In addition, a random distribution of points with the same selection 
function as the cluster data is used to model the effects 
of the survey geometry on the recovered power spectrum. 
The measured power spectrum is sensitive to the level of smoothing applied 
to the selection function, though the tests presented 
by Tadros \etal (1998) suggest that the turnover found in the power spectrum 
is robust to changes in the smoothing.
The Gazta\~{n}aga \etal (1995) 
approach does not require an estimate of the survey 
selection function (see Efstathiou \etal 1990), but does assume that the 
fluctuations are Gaussian. However, relaxing this assumption does not 
change the results significantly.
Furthermore, the estimators used for the power spectrum and 
for the variance of counts in cells could respond in different ways 
to uncertainties in the mean density of clusters and to the irregular 
boundary of the APM survey.

\subsection{Model parameters}
\label{s:model}

We now compare theoretical predictions for 
the power spectrum of clusters in redshift space 
with measurements of the power spectrum of the APM cluster redshift survey. 
We make a number of assumptions in our model: 

\begin{itemize}
\item[(i)] 
We assume that primordial density fluctuations are Gaussian.
The bulk of the available evidence suggests that any non-Gaussianity 
found in the distribution of large scale structure today is consistent 
with the gravitational evolution of an initially 
Gaussian density field (Moore \etal 1992; Gazta\~{n}aga 1994; 
Canavezes \etal 1998; Hoyle, Szapudi \& Baugh 2000; Szapudi \etal 2000). 
Relaxing this assumption would affect the model predictions for 
the abundance of clusters and therefore, through changing the bias factor, 
the clustering of clusters (Robinson, Gawiser \& Silk 2000).

\item[(ii)]
The dark matter in the universe is assumed to be cold dark matter.

\item[(iii)]
We assume that the shape of the dark matter power spectrum is the 
same as that measured for the galaxy power spectrum on 
the scales that we consider.
Local biasing schemes yield an asymptotically constant bias on large scales 
between fluctuations in the galaxy and mass distributions  
(Coles 1993; see Cole \etal 1998 for realisations of this process). 
Observational support for this assumption is derived from the comparison of 
the power spectra of different types of object. Peacock \& Dodds (1994) 
found that on large scales, the power spectra of optical galaxies, 
radio galaxies and clusters are consistent with a single underlying 
power spectrum after applying different, constant relative bias factors to the 
individual measurements.
Tadros \etal (1998) also demonstrate that the power spectra of APM clusters 
and galaxies have the same shape on large scales and are related by 
a constant relative bias factor.
The connection with the shape of the dark matter spectrum is made 
theoretically. Colberg \etal (2000) show that in real space the power 
spectrum of clusters is the same shape as the power spectrum of the 
underlying dark matter. In Section \ref{s:pk} we have demonstrated that 
the same conclusions hold in redshift space, when cluster redshift errors 
are included.

The shape of the real space power spectrum of galaxies is well determined 
(Maddox \etal 1990; Baugh \& Efstathiou 1993, 1994; Peacock \& Dodds 1996; 
Gazta\~{n}aga \& Baugh 1998). 
We adopt values of $\Gamma=0.2, 0.25$, consistent with the 
recent determinations by Efstathiou \& Moody (2001) and 
Eisenstein \& Zaldarriaga (2001).

\item[(iv)] 
We apply the methodology outlined in Section \ref{s:theory} and 
assume that the construction of a cluster sample specified by a 
space density is equivalent to taking all clusters above some mass threshold. 
To compare with sample B drawn from the APM cluster survey, 
we adopt a value of $d_c = 30.9 h^{-1}$Mpc.
\end{itemize}

Specifically, our model has two parameters: 

\begin{itemize}
\item[(i)] The cosmological density parameter, $\Omega_{0}$. 
We consider values in the range $\Omega_{0}=0.01-1$,  
with the condition that the model universe is spatially flat, 
i.e. if $\Omega < 1$, then we adopt a cosmological constant, 
$\Lambda$, such that $\Omega + \Lambda c^{2}/(3 H^{2}) = 1$. 
This is in agreement with the location of the first Doppler peak 
reported by the BoomeranG and Maxima cosmic microwave background 
experiments (Balbi \etal 2000; de Bernardis \etal 2000). 

\item[(ii)] The amplitude of density fluctuations, as specified by 
the linear {\it rms} variance in spheres of radius $8h^{-1}$Mpc, $\sigma_{8}$.

\end{itemize}
The theoretical models are computed for a mean cluster 
redshift of $z=0.083$, which is appropriate for APM sample B.

When comparing the model predictions with the data, 
we use the APM power spectrum results obtained for a 
cosmology with $\Omega=1$. Tadros \etal (1998) also 
compute the power spectrum of APM clusters in a 
background cosmology defined by $\Omega_{0}=0.2$, 
${\Lambda_{0}c^{2}}/(3H^{2}_{0}) = 0.8$, 
and recover the same shape of power spectrum, but with 
an amplitude that is approximately $25\%$ higher. 
Therefore we make a systematic error when comparing data 
derived assuming $\Omega=1$ with a model in which  
a different value of $\Omega$ is adopted.
To partially compensate for this, we take a pessimistic view 
of the errors on our theoretical predictions, adopting the 
level of the discrepancy found between the model and 
the $\Lambda$CDM simulation results of $15\%$.

\subsection{Results}
\label{s:res2}

The theoretical models are assessed by computing a 
figure of merit: 
\begin{equation}
\chi^{2} = \sum  
\left(P_{\rm model}(k) - P_{\rm data}(k)\right)^{2} /
\left( \sigma^{2}_{\rm data}(k) + \sigma^{2}_{\rm model}(k) \right )
\end{equation}
where the sum is over all the data points plotted in 
Fig \ref{fig:pkdata}(b) and we take 
$\sigma_{\rm model}(k) = 0.15 P_{\rm model}(k)$ 
(see Robinson 2000).
We ignore any covariance between measurements of the power spectrum 
at different wavenumbers.

\begin{figure*}
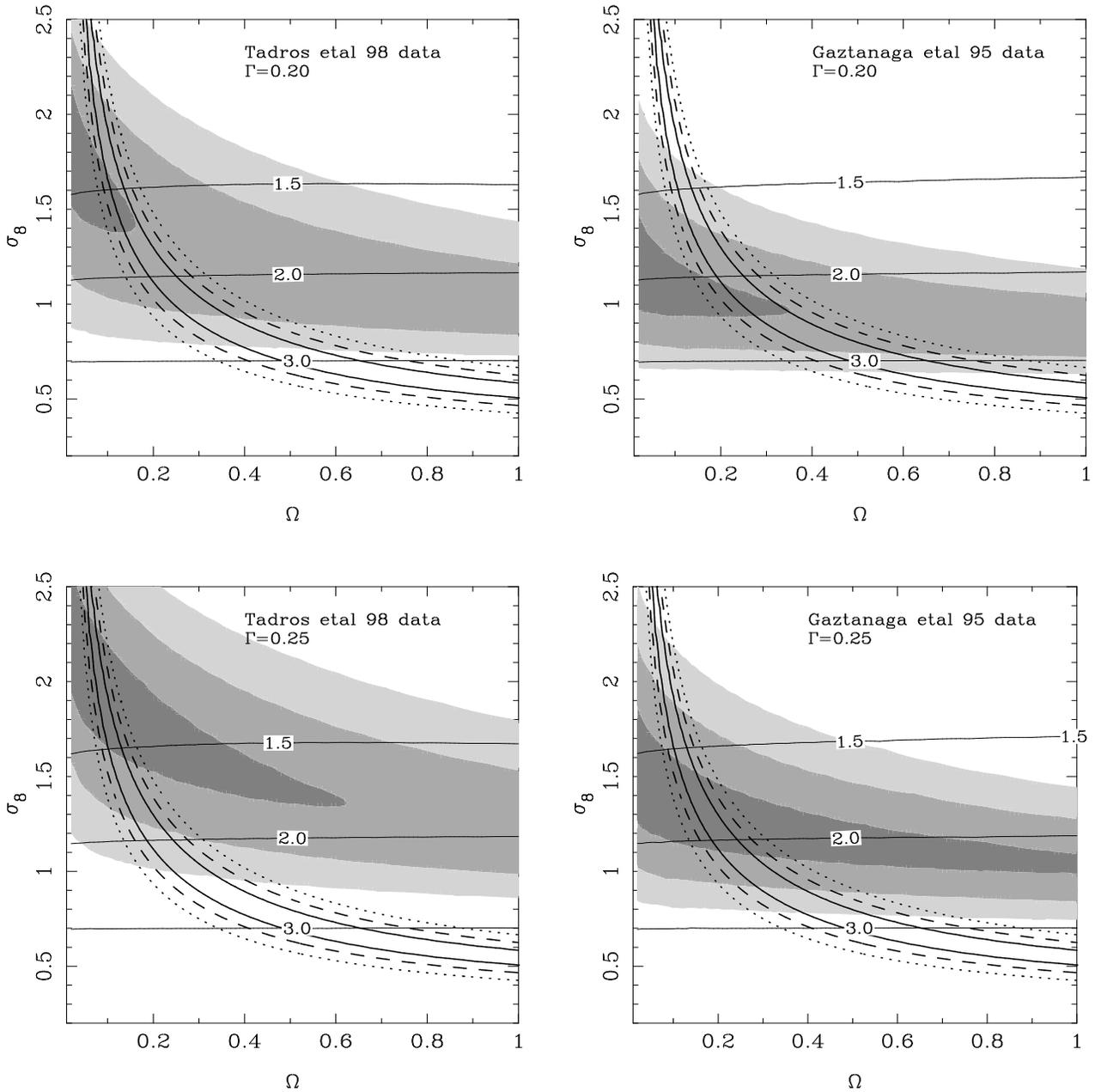

\begin{picture}(400,500)
\put(-40,250){\psfig{file=tadros.g0.2.ps,width=8cm,height=8cm,angle=-90}}
\put(210,250){\psfig{file=gazta.g0.2.ps,width=8cm,height=8cm,angle=-90}}
\put(-40,0){\psfig{file=tadros.g0.25.ps,width=8cm,height=8cm,angle=-90}}
\put(210,0){\psfig{file=gazta.g0.25.ps,width=8cm,height=8cm,angle=-90}}
\end{picture}
\caption{
The constraints on the model parameters $\sigma_{8}$ and $\Omega_{0}$ 
from APM cluster power spectrum data. 
The shaded regions show contours of $\Delta \chi^{2}=1,4 $ and $9$.
The almost horizontal lines show parameter combinations for which the 
effective cluster bias is $b_{\rm eff}=1.5,2$ and $3$.
The left hand panels show the outcome of comparing the models to the Tadros 
\etal (1998) power spectrum data, the right hand panels are obtained using the 
power spectrum inferred from the Gazta\~{n}aga \etal (1995) variance of 
counts in cells. 
The upper row is for models with $\Gamma=0.2$, the lower row for 
$\Gamma=0.25$.
The lines show $\Delta \chi^{2}$ contours for the constraint on 
$\sigma_{8}$ and $\Omega_{0}$ from the local abundance of hot X-ray clusters 
and are reproduced in each panel (Eke, Cole \& Frenk 1996b).
}
\label{fig:chi}
\end{figure*}

The results are shown in Fig. \ref{fig:chi}. The shaded regions 
show models that are within $\Delta \chi^{2} < 1, 4, 9$ of the best 
fitting model when compared to the cluster power spectrum data indicated 
in the legend on each panel.
The legend also gives the value of the power spectrum shape parameter, 
$\Gamma$, used in the models.
In Fig. \ref{fig:chi}, the left hand panels show the results of the comparison 
to the Tadros \etal (1998) power spectrum data, and the right hand panels 
show the results using the power spectrum inferred from the Gazta\~{n}aga 
\etal (1995) measurement of the variance of APM clusters.
The top row of the Fig. is for models with $\Gamma = 0.2$, the bottom 
row is for $\Gamma=0.25$.
We also plot the same $\Delta \chi^{2}$ contours for the constraint 
on the value of $\sigma_{8}$ derived from the abundance of hot X-ray 
clusters in the local universe by  Eke, Cole \& Frenk (1996b) 
(shown by the solid, dashed and dotted lines). 
The error on $\sigma_{8}$ quoted by these authors is $8\%$. 
The Eke \etal results agree  with those from a recent reanalysis 
of the current cluster X-ray temperature data and of the theoretical 
framework of this constraint carried out by Pierpaoli, Scott \& White (2000). 
The almost horizontal lines show the models in which clusters with a mean 
separation of $d_{c} = 30.9 h^{-1}$Mpc have effective bias factors of 
$b_{\rm eff}=1.5,2$ and $3$.

The $\Delta \chi^{2}$ contours derived from the cluster power spectrum 
constraint are almost parallel to the $\Omega$ axis, particularly for 
$\Omega > 0.2$. Thus the amplitude of the cluster power spectrum in the 
models is determined largely by the value of $\sigma_{8}$ 
(see Mo, Jing \& White 1996).
The best fitting models with $\Gamma=0.25$ have a lower $\chi^2$ per degree 
of freedom than the $\Gamma=0.2$ models.  
The difference in the amplitude of the Tadros \etal (1998) and 
Gazta\~{n}aga \etal (1995) data is readily apparent from the shift in the 
location of the $\Delta \chi^2$ contours. 
The models that come closest to matching the Gazta\~{n}aga \etal (1995) 
data tend 
to have an effective cluster bias factor of $b_{\rm eff} \sim 2$, whilst 
the best fitting models for the Tadros \etal (1998) data tend to 
have $b_{\rm eff} \sim 1.5$.

The constraints on model parameters derived from the power spectrum 
data are much broader than those obtained from the abundance of 
rich clusters. Moreover, the shapes of the two sets of contours 
are different. A rough fit to the middle of the $\Delta \chi^{2} < 1$ 
contour for the Gazta\~{n}aga \etal (1995) 
data, using models with $\Gamma=0.25$, yields 
$\sigma_{8} = \Omega_{0}^{-0.18 + 0.07 (1 - \Omega_{0})}$, whereas the 
cluster abundance constraint scaling is approximately 
$\sigma_{8} \propto \Omega^{-0.56}_{0}$.

\begin{figure}
\centerline{\psfig{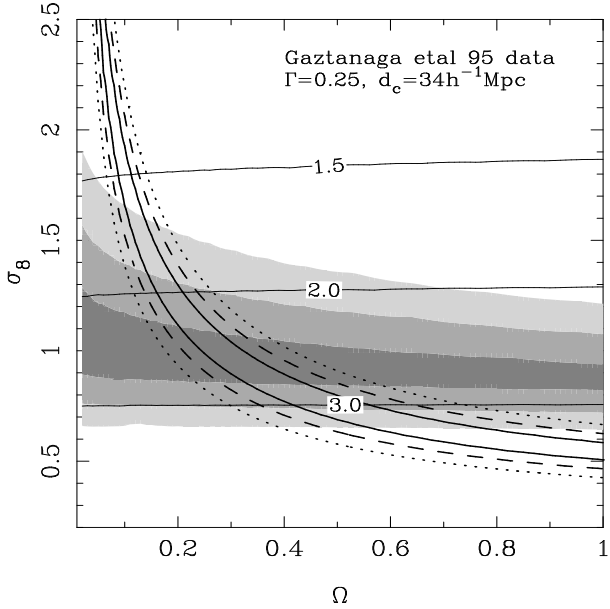}}
\caption{
The constraints on model parameters when the abundance of APM clusters 
is reduced by $30\%$, so that $d_{c}=34h^{-1}$Mpc. The lines and shading 
are the same as for Fig \ref{fig:chi}.
}
\label{fig:chi2}
\end{figure}

We illustrate the consequences of an error in the space density 
of APM clusters in Fig. \ref{fig:chi2}. The estimated space density 
is somewhat sensitive to the way in which the cluster selection 
function is normalised (Efstathiou \etal 1992). 
Furthermore, any projection effects that 
persist in the machine constructed catalogues could be responsible for 
moving poorer clusters into the sample. We therefore plot 
in Fig. \ref{fig:chi2} the constraints on model parameters 
assuming that $d_{c}=34h^{-1}$Mpc; 
this represents a $10\%$ error in $d_c$, corresponding to the space 
density of APM clusters being $30\%$ lower than assumed in Fig. \ref{fig:chi}.
When the space density of the sample decreases, the minimum mass threshold 
that defines the sample increases, leading to larger effective bias 
parameters. The $\Delta \chi^{2}$ contours therefore shift down to 
lower values of $\sigma_{8}$.

\begin{figure}
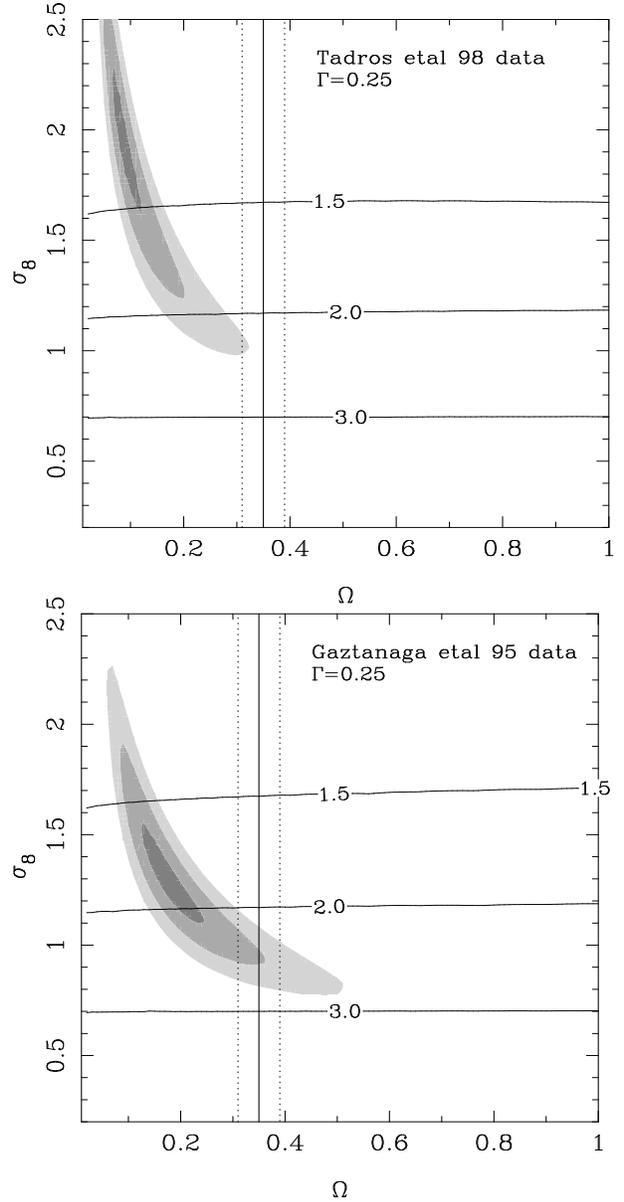

\begin{picture}(300,450)
\put(0,225){\psfig{file=tadros.g0.25.comb.ps,width=8cm,height=8cm,angle=-90}}
\put(0,0){\psfig{file=gazta.g0.25.comb.ps,width=8cm,height=8cm,angle=-90}}
\end{picture}
\caption{
The best fitting model parameters when the constraints from the 
cluster power spectrum and abundance measurements are combined. 
The contours show $\Delta( \chi^{2}_{P(k)} + \chi^{2}_{abun}) =1, 4$ and $9$.
The top panel shows the $\Delta \chi^{2}$ contours when the Tadros \etal 
(1995) $P(k)$ is used and the bottom panel is for the 
Gazta\~{n}aga \etal (1995) data.
A model power spectrum shape of $\Gamma=0.25$ is adopted and the model 
clusters are defined by $d_{c}=30.9h^{-1}$Mpc. 
The vertical line shows the value of $\Omega$ that is consistent with the 
adopted value of $\Gamma$, the CDM definition of $\Gamma=\Omega h$ and the 
measurement of Hubble's constant by Freedman \etal (2001). 
The vertical dotted lines show the range of $\Omega$ permitted within 
the $1\sigma$ errors quoted on Hubble's constant.
}
\label{fig:chi3}
\end{figure}

Finally, we consider the constraints on the model parameters that 
are obtained when the two datasets, the cluster power spectrum  
and local abundance of hot X-ray clusters, are combined. 
In Fig. \ref{fig:chi3},  we plot the $\Delta \chi^2$ contours after 
adding the $\chi^2$ values from comparing the model to the power 
spectrum data and to the observed cluster abundance. 
This operation assumes that the two datasets are independent. 
Due to the smaller errors, the cluster abundance constraint 
has the largest influence on the resulting $\chi^2$ contours. 
The vertical line on each panel shows the value of $\Omega$ that is 
consistent with the chosen $\Gamma$, given the recent determination of 
Hubble's constant as $H_{0} = 72 \pm 8 {\rm kms}^{-1}{\rm Mpc}^{-1}$ by 
Freedman \etal (2001), and assuming that $\Gamma = \Omega h$, where 
$H_{0}=100 h {\rm kms}^{-1}{\rm Mpc}^{-1}$. The dotted lines indicate the 
range of values of $\Omega$ allowed following this prescription when 
the $1 \sigma$ errors on Hubble's constant are taken into account.
The combined dataset favours low values of $\Omega$, with 
$\sigma_{8}=1$--$1.5$ for the Gazta\~{n}aga \etal measurement and 
$\sigma_{8}=1.6$--$2.3$ for the Tadros \etal power spectrum.
In both cases, the shape of the power spectrum expected in the 
most easily motivated cold dark matter model, given the best fitting 
value of $\Omega$ is somewhat discrepant with the shape that we 
have assumed for the power spectrum; this disagreement is significant 
for the Tadros \etal data.

\section{Discussion}
\label{s:end}

The main goal of this paper was to establish the degree of distortion  
expected in the clustering pattern of rich clusters when viewed in 
redshift space.
For this purpose, we undertook high precision measurements of the 
clustering of massive dark matter haloes, with the same space density 
as observational samples of clusters, using the largest extant 
cosmological simulations, the Virgo Consortium's Hubble Volume 
simulations. 

The {\it rms} peculiar motions of clusters in the Hubble Volume 
simulations are modest, of the order $300$--$350 {\rm kms}^{-1}$ for a 
single cluster. Examples of pairwise velocity dispersions 
of the order of several thousand kilometers per second are found in 
the simulations, but are relatively rare. 
We find no evidence for {\it any} enhancement of the correlation amplitude 
along the line of sight arising from peculiar motions.
The main type of distortion to the clustering pattern in redshift 
space is a flattening of the contours of $\xi(\sigma,\pi)$ due to 
coherent motions of clusters. 
The shape of the $\xi(\sigma,\pi)=1$ contour is remarkably similar to 
that found by Miller \etal (1999) for a recent survey of $R \ge 1$ Abell 
clusters (see their Fig. 7; we note however that lower amplitude 
contours in their Fig. still show some enhancement along the line of sight).  
If a redshift error is assigned to each cluster in the simulation 
(with a single cluster {\it rms} of $500 {\rm kms}^{-1}$), 
then we reproduce the form of the distortion seen in measurements 
of $\xi(\sigma,\pi)$ from the APM cluster redshift survey, 
confirming that this survey is largely free from projection effects.

Similar conclusions were reached in a study of the redshift space distortions 
in the clustering of galaxies and groups in the Updated Zwicky Catalogue 
by Padilla \etal (2001). In this case, the groups are identified in three 
dimensions rather than in projection. A different set of issues need to be 
addressed when tuning a three dimensional group-finding algorithm, but 
superposition of groups along the line of sight should only occur for 
systems with high internal velocity dispersions. 
Padilla \etal (2001) found a strong enhancement 
in the line of sight clustering 
for galaxies. No such feature was detected in the correlation function 
of groups. These results were compared with predictions from N-body 
simulations, and the same form of distortion was found.

The analytic model outlined in Section \ref{s:theory} provides 
an accurate description of the redshift space power spectrum of clusters 
measured in Hubble Volume simulations, when errors in the cluster 
redshifts are included.
We then compared the model predictions to the power spectrum measured 
from the APM cluster redshift survey in order to constrain the 
cosmological parameters $\Omega$ and $\sigma_{8}$.
This is a more direct approach than a comparison with the cluster 
correlation length and thus avoids uncertainties regarding the way in which 
the correlation length is derived from the measured correlation function. 

We have chosen to focus our attention on the APM cluster redshift survey, 
as this is the largest volume survey available that has been 
constructed from survey plates using a well specified, automated procedure. 
Miller \& Batuski (2001) measure the power spectrum of 
a sample of $R\ge 1$ Abell clusters that does not display large enhancements 
in the line of sight clustering and which covers a larger volume 
than the APM survey. These authors find no evidence for a turnover 
in the cluster power spectrum and probe larger scales than the 
measurements of Gazta\~{n}aga \etal (1995) and Tadros \etal (1998).
However, the space density of clusters in Miller \& Batuski's sample 
changes by a factor of two in a northern extension of the survey, 
which contributes much of the signal to the power spectrum 
measurement on large scales.
Therefore more detailed modelling of the observational sample is 
required than we have attempted in this paper in order to make a 
realistic comparison with these data. 
Furthermore, the integrity of the Abell catalogue remains open 
to question, even for $R\ge1$ clusters. 
Van Haarlem, Frenk \& White (1997) demonstrated that mock Abell 
cluster catalogues constructed from N-body simulations 
suffer from significant incompleteness for $R\ge1$ clusters.  
Power spectrum measurements have also been made using X-ray selected 
samples of clusters (e.g. Zandivarez, Abadi \& Lambas 2001, 
Schuecker \etal 2001). 
Such samples have the appeal of greatly reducing projection effects as the 
X-ray emission is dominated by the high density core of a cluster's 
dark matter halo. 
However, flux limited X-ray surveys mix clusters of different richness, 
so more careful modelling of the cluster selection is required to make 
robust theoretical predictions (Borgani \etal 1999; Moscardini \etal 2000).

We assume that the dark matter power spectrum has the shape measured 
for the galaxy power spectrum on large scales. This is a reasonable 
approximation if galaxy formation is a 
local process (Coles 1993; Cole \etal 1998). 
The model parameters that we vary to fit the cluster power spectrum data 
are then the cosmological density parameter $\Omega$ and the fluctuation 
amplitude $\sigma_{8}$. 
Within this scheme, the cluster power spectrum does not provide any 
constraint on $\Omega$ (see Mo, Jing \& White 1996). 
The constraints on $\sigma_{8}$ are fairly broad as a result 
of the relatively large errors on the measured cluster power spectrum.
This situation can be improved if additional information is used. 
Tadros \etal (1998) compare the power spectrum of galaxies and clusters 
in the APM survey and find that they have the same shape and can be 
related by a relative bias of $b \approx 1.8$. If one assumes that 
APM galaxies are essentially unbiased tracers of the dark matter on 
large scales (Gazta\~{n}aga 1994), then this additional constraint 
can be used to exclude models in which APM-like clusters 
have an effective bias that is very different from $b \sim 1.8$. 
However, we caution against the use of measurements of clustering 
in redshift space to restrict the value of an effective bias parameter 
in real space (see Fig. \ref{fig:pkb}).

The range of acceptable model parameters is tightened considerably if 
the cluster power spectrum data is combined with measurements of the 
local abundance of rich clusters. 
The cluster abundance measurements themselves constrain the parameter 
combination $\sigma_{8} \Omega^{0.6}$ (White, Efstathiou \& Frenk 1993). 
The constraint on these parameters from the power spectrum data has 
a different dependence on $\Omega$, thus allowing the degeneracy 
to be lifted.
The best model parameters obtained using the combined data sets are 
$\sigma_{8} \approx 1.25$ and $\Omega \approx 0.2$.
The best fitting value of $\Omega$ is inconsistent with the value suggested 
by our choice of $\Gamma$, if we use the most readily motivated prescription 
for setting the shape of the power spectrum in cold dark matter models, 
$\Gamma=\Omega h$, and adopt a recent measurement of Hubble's constant by 
Freedman \etal (2001). The discrepancy is severest for the 
Tadros \etal (1995) power spectrum data (see the discussion of this 
data in Gawiser \& Silk 1998).
Whilst values of the shape parameter $\Gamma < \Omega h$ can be motivated 
physically by postulating a large baryon fraction or 
decaying neutrinos (Eisenstein \& Hu 1998 ; White, Gelmini \& Silk 1995), 
it is difficult to see how a value for $\Gamma$ that is much larger than 
$\Omega h$ could be accommodated.
However, this situation could be alleviated if the abundance of 
APM clusters has been overestimated and a larger value of $d_c$ 
turns out to be more appropriate.

The cluster power spectrum data will improve significantly in the 
near future upon completion of the 2dF and SDSS galaxy redshift surveys 
(Colless 1999; York \etal 2000). 
Group catalogues will be constructed in three dimensions and the large 
volumes covered by the surveys mean that these catalogues will contain 
large numbers of clusters. 
Moreover, a comparison between the properties of clusters selected in 
projection and in redshift space will permit fine tuning of the 
two dimensional algorithms, that can then be applied to the deeper 
parent photometric catalogues.

\section*{Acknowledgments}
This work  was supported in part by a British Council grant 
for exchanges between Durham and Cordoba, 
by CONICET, Argentina and  by a PPARC rolling grant at the University 
of Durham.  
CMB acknowledges receipt of a Royal Society University Research Fellowship.
We acknowledge the Virgo Consortium for making the Hubble Volume 
simulation output available and  thank Adrian Jenkins for helpful 
discussions and for assistance in accessing and using the cluster catalogues.
We are indebted to Shaun Cole for supplying us with a copy of his 
code to compute power spectra on high resolution grids and for a critical 
reading of the manuscript.

\end{document}